\documentclass[epj,final]{svjour}
\usepackage{epsfig}
\usepackage{amsmath,amssymb}

\begin{document}
\title{Phase diagram of the three-dimensional Hubbard model at half filling}
\author{R. Staudt\inst{1}, M. Dzierzawa\inst{1}, and A. Muramatsu\inst{2}} 
\institute{Institute of Physics, University of Augsburg, 
  Universit\"atsstra{\ss}e 1, D-86135 Augsburg \and
  Institute of Physics, University of Stuttgart, 
  Pfaffenwaldring 57, D-70550 Stuttgart, Germany}
\date{Received: date / Revised version: date}

\abstract{
We investigate the phase diagram of the three-dimensional Hubbard model
at half filling using quantum Monte Carlo (QMC) simulations.
The antiferromagnetic N\'eel temperature $T_N$ is determined from the
specific heat maximum in combination with finite-size scaling of the magnetic 
structure factor.
Our results interpolate smoothly between the asymptotic solutions for
weak and strong coupling, respectively, in contrast to previous QMC simulations. 
The location of the metal-insulator transition in the 
paramagnetic phase above $T_N$ is
determined using the electronic compressibility as criterion.
}
\PACS{
     {71.10.Fd}{} \and
     {71.10.Hf}{} \and
     {71.30.+h}{} 
     }
\authorrunning{R. Staudt {\it et al.} }
\titlerunning{Phase diagram of the 3D Hubbard model at half filling}
\maketitle

\section{Introduction}

The canonical lattice model for correlated electrons is
the Hubbard model \cite{Gut63,Hub63,Kan63} defined by the hamiltonian
\begin{equation}
\hat H = -t \sum_{<ij>,\sigma}(\hat{c}^{\dagger}_{i\sigma}\hat{c}^{ 
}_{j\sigma}+\hat{c}^{\dagger}_{j\sigma}\hat{c}^{ }_{i\sigma})
 + U\sum_{i}\hat{n}_{i\uparrow}\hat{n}_{i\downarrow}
\end{equation}
where $\hat{c}^{\dagger}_{i\sigma}\;(\hat{c}^{ }_{i\sigma})$ are electron 
creation (annihilation) operators, $\langle ij \rangle$ denotes a pair of
neighboring lattice sites and $t$ and $U$ are the hopping matrix element
and the onsite Coulomb energy, respectively.
At half filling the ground state of the 3D Hubbard model on a simple cubic lattice
has antiferromagnetic long range order for all positive values of $U$ due to
the perfect nesting of the Fermi surface.
At finite temperatures the sublattice magnetization is reduced
by thermal fluctuations and a transition to a paramagnetic phase occurs
at the N\'eel temperature $T_N$. 
In the strong coupling limit the energy scale for magnetism is obtained
by mapping the Hubbard model to the antiferromagnetic spin $1/2$
Heisenberg model with exchange coupling $J = 4\, t^2/U$.
For small $U$ antiferromagnetism results from the Fermi surface instability and
the relevant energy scale is set by the BCS like expression
$T_N \propto t \exp(-1/(\rho_0 U))$ 
where $\rho_0$ is the density of states
at the band center.

There have been many attempts to calculate  
$T_N$ over the whole range of $U$ using both 
QMC simulations \cite{Hir87,Sca89} and analytical
approaches including variational methods \cite{Kak8586}, 
Hartree Fock theory \cite{Don9194}, 
strong coupling expansions \cite{Szc95},
dynamical mean field theory (DMFT) \cite{Jar92,Geo93,Ulm95},
the two-particle self-consistent formalism
\cite{Dar00} and spin fluctuation theory \cite{Sin98}.
A critical comparison of the various approaches can be found in refs. \cite{Szc95} and
\cite{Dar00}. A common drawback of mean-field like approximations is 
that in the strong coupling limit they fail to 
reproduce the correct Heisenberg result $T_N = 3.83 \,t^2 / U$ \cite{Rus74}
and instead reduce to the Weiss molecular field theory with 
$T_N^{mf} = 6 \,t^2 / U$. Strong coupling expansions, on the other hand,
break down in the Fermi surface instability regime.
As to numerical methods, the results of previous QMC simulations \cite{Hir87,Sca89}
do not agree with each other and have 
been questioned by Hasegawa \cite{Has89} who argued that they overestimate
$T_N$ considerably. Nevertheless, for lack of alternatives
the QMC results of refs. \cite{Hir87,Sca89} have served as benchmarks for
analytical approaches over the last ten years despite the controversy
concerning their reliability. Clearly, there is need for improved
QMC simulations of the 3D Hubbard model with better statistics and for 
larger systems than previously accessible.

Besides antiferromagnetism the second important phenomenon described by the half-filled 
Hubbard model is the interaction-induced metal-insulator transition,
known as Mott-Hubbard transition \cite{Mot74}. This transition occurs when the ratio
of the interaction strength $U$ and the bandwidth $W$ exceeds
some critical value of order one. Unfortunately, the presence of antiferromagnetic
order makes it impossible to observe the Mott-Hubbard transition
in the 3D Hubbard model at temperatures below $T_N$. 
DMFT calculations for the fully frustrated Hubbard model \cite{Geo96} where
antiferromagnetism is completely suppressed
predict a first order metal-insulator transition line that persists up to a critical
point at $(U_c,T_c)$ in the phase diagram, followed by
a crossover region above this critical point.
In this paper we examine if such a first order transition line
can also be observed in the 3D Hubbard model or if it is completely occluded
by the antiferromagnetic phase below $T_N$.

\section{Magnetic phase transition}

We have studied the 3D Hubbard model on a simple cubic lattice with periodic
boundary conditions
using a finite temperature, grand canonical 
QMC algorithm \cite{Bla81,Hir82} which is based on a discrete Hubbard-Stratonovich
decoupling of the Hubbard interaction term. In this algorithm, the inverse temperature 
or imaginary time $\beta$  
has to be divided into a finite number of steps $\Delta \tau$ which
introduces an error $\propto \Delta \tau^2 t U$ into the calculations. 
We have chosen $\Delta \tau^2 t U = 0.1$ after making sure
that the results are not significantly affected
by the extrapolation  $\Delta \tau \rightarrow 0$. 
A typical simulation on a $L\times L\times L$ lattice
consisted of between 100000 (L = 4) and 2000 (L = 10) measurement
sweeps that were grouped into 20 blocks in order to estimate the statistical error
of the QMC data. 
QMC simulations frequently suffer from the minus sign problem, i.e. the
fact that the fermionic determinant that serves as probability weight function
is not always positive, in particular at low temperatures. 
At half filling where most of our simulations were performed this problem does not
exist due to particle-hole symmetry,
but even in the calculations of the compressibility that require simulations
away from half filling we never encountered any serious 
minus sign problem since we worked at relatively high temperatures. 

\begin{figure}[tb]
\centerline{\psfig{file=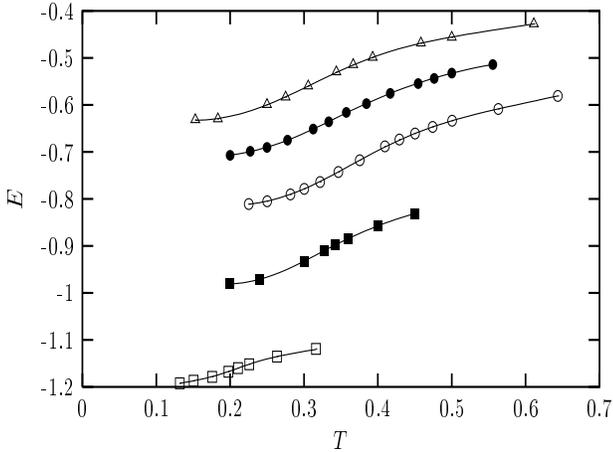,width=8cm,height=6cm,angle=0}}
\caption{Energy as function of temperature for $L = 4$ and $U = 4, 6, 8,
10, 12$ (from bottom to top). Error bars are much smaller than the 
size of the symbols.
The curves are obtained with the procedure described in the text.}
\label{eoft}
\end{figure}

\begin{figure}[tb]
\centerline{\psfig{file=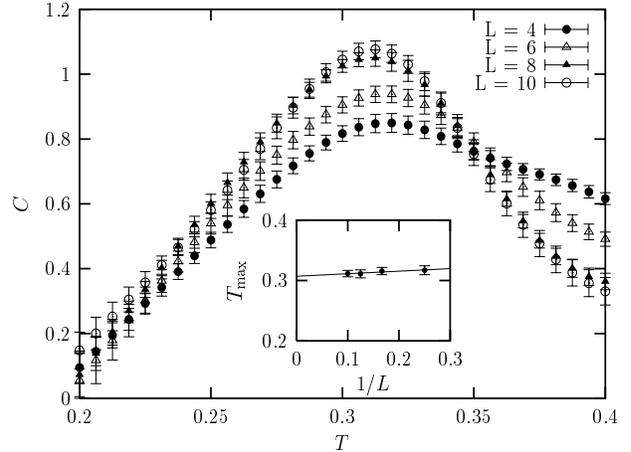,width=8cm,height=6cm,angle=0}}
\caption{Specific heat as function of temperature for $U = 6$ and $L = 4,6,8,10$.
The inset shows the position of the maximum of the specific heat vs. $1/L$.}
\label{cv}
\end{figure}

Our first goal is to determine the antiferromagnetic N\'eel temperature as function
of the interaction strength $U$. To this end we have calculated
the specific heat which is the central quantity to characterize thermodynamical properties
of many-particle systems and is easily accessible both experimentally and from numerical
simulations. Since phase transitions are usually accompanied by a maximum or
a divergence of the specific heat the transition temperature can be
determined from specific heat data without measuring 
the order parameter directly. Of course, not every maximum of the specific heat 
is associated with
a phase transition; one has to verify that the order parameter indeed
changes from zero to a nonzero value at the presumed transition temperature. 
In the Hubbard model for sufficiently large $U$ there are two maxima of the specific heat,
one at low temperatures $\propto t^2/U$ reflecting spin excitations, the second on a
higher energy scale $\propto U$ associated with charge excitations.
Here we concentrate on the low temperature peak which we use to determine the
N\'eel temperature.

The specific heat can either be calculated via $C = \partial E / \partial T$ or 
from the
energy fluctuations, $C \propto \langle \hat H^2 \rangle -  \langle \hat H \rangle^2$.
We employ the first method which turns out to be
numerically more accurate. The following procedure is used:
First the energy is calculated with QMC in a temperature range 
where the phase transition is expected to take place.
The simulations are performed at fixed imaginary time slice 
$\Delta \tau$ for all temperatures.
An extrapolation $\Delta \rightarrow 0$ is not necessary since the finite $\Delta \tau$ 
corrections to the energy are nearly temperature independent over the 
relatively small temperature
range under consideration and even taking into account corrections to linear order in $T$ 
does not affect the position of the specific heat maximum.
In order to calculate the specific heat from the energy data 
the temperature interval is divided in a number of equidistant
values. For each of these temperatures $T_i$ the value of the specific heat
$C_i$ is determined such, that {\em i)} the resulting curve is as smooth as possible
and {\em ii)} the energies $E_{\makebox{fit}}$
obtained by numerical 
integration of the specific heat values $C_i$ are as close as possible to the QMC data.
In practice this is achieved
by minimizing the quantity $\lambda_1 A + \lambda_2 B$ where
$A = \sum_i (C_{i+1} - 2 C_i + C_{i-1})^2$ 
controls the smoothness of the fit while
$B = \sum ((E_{\makebox{fit}} - E_{\makebox{QMC}})/
\sigma_{\makebox{QMC}})^2$ measures the
deviation between the fit and the QMC data.
This procedure avoids to chose a specific functional form for the fit function
which could bias the results in some way or the other.
The ratio of the parameters $\lambda_1$ and $\lambda_2$ is fixed such that
the energy values obtained by the minimization procedure lie within the statistical errors
$\sigma_{\makebox{QMC}}$ of the QMC data.
We have checked that the final results are very robust against moderate
variations of the ratio $\lambda_1 / \lambda_2$.
In order to estimate the error bars for the specific heat we have performed an average 
over many different input energy data sets 
obtained by adding random noise of the order of
$\sigma_{\makebox{QMC}}$ to the mean values
$E_{\makebox{QMC}}$.
Of course, any noise-reducing algorithm tends to wash out 
sharp structures like cusps or discontinuities.
On the other hand, the specific heat curves obtained
from extensive QMC simulations of the three-dimensional
spin 1/2 Heisenberg model \cite{San98} are as smooth as
ours indicating that there exist no sharp structures that could
be lost by the procedure that we use to calculate
the specific heat, at least for the system sizes that we consider.
In any case, the location of the maximum which we are mainly interested in
is only very little affected.

Fig. \ref{eoft} shows the energy as function of temperature for $L = 4$ and several
values of $U$. Here and in the following figures we have fixed the energy scale by
setting $t = 1$.
The curves connecting the data are obtained with the fitting
procedure described above. They all have a point of inflection indicating
that the specific heat maxima are contained in the respective temperature intervals.
In Fig. \ref{cv} the specific heat for $U = 6$ is displayed as function of temperature for
$L = 4,6,8$ and $10$. The maximum value of $C$ increases somewhat 
with increasing system size while simultaneously the position of the maximum 
$T_{max}$ is slightly shifted to lower temperatures. 
There is no indication of a divergence for large systems. This 
is in agreement with the assumption that the half-filled
Hubbard model belongs to the 3D Heisenberg universality class 
with a negative exponent
$\alpha \approx -0.11$ which means that the specific heat
has only a cusp but no divergence at the critical temperature.
High precision QMC simulations of the 3D
spin 1/2 Heisenberg model \cite{San98}
confirm this behavior. In the Heisenberg model the shift of the maximum of $C$ 
between $L = 4$ and the infinite lattice is about five percent. 
In the inset of Fig. \ref{cv} 
the peak temperature $T_{\makebox{max}}$ is plotted vs $1 / L$ indicating that
in the Hubbard model finite-size corrections are quite small as well.
A linear fit yields $T_N = 0.31 \pm 0.01$.


\begin{figure}[tb]
\centerline{\psfig{file=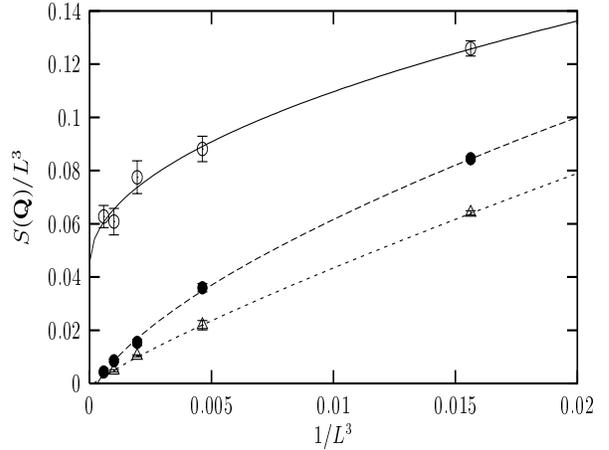,width=8cm,height=6cm,angle=0}}
\caption{Finite size scaling of the magnetic structure factor 
for $U = 6$ and $T = 0.3, 0.36$ and $0.4$ (from top to bottom).}
\label{su6}
\end{figure}

We now demonstrate that the peaks in the specific heat are indeed associated
with the antiferromagnetic phase transition. To this end we have calculated
the magnetic structure factor
\begin{equation}
S({\bf Q}) = \frac{1}{L^3} \sum_{i,j} \, {\rm e}^{i \,{\bf Q}\,({\bf R}_i-{\bf R}_j)}
 \, \langle (\hat n_{i\uparrow} - \hat n_{i\downarrow})
            (\hat n_{j\uparrow} - \hat n_{j\downarrow})
\rangle
\end{equation}
where ${\bf Q} = (\pi,\pi,\pi)$ is the antiferromagnetic wave vector.
$S({\bf Q})$ is related to the sublattice magnetization $m$ via
\begin{equation}
\frac{S({\bf Q})}{L^3} = m^2 + f(L)
\end{equation}
where $f(L) \rightarrow 0$ for $L \rightarrow \infty$. In order to extrapolate
the finite lattice QMC data to the thermodynamic limit one has to know the
asymptotic behavior of $f(L)$ for large $L$.
Right at the critical temperature the structure factor $S({\bf Q})$
should scale with the system
size as $L^{2-\eta}$ where $\eta \approx 0.03$ for the 3D Heisenberg universality
class \cite{San98} and therefore $f(L) \propto L^{-1.03}$. 
In the ordered phase at low temperatures, spin wave theory predicts 
$f(L) \propto L^{-2}$ assuming
a linear magnon dispersion. On the other hand,  in the paramagnetic phase above $T_N$
spin correlations decay exponentially and we expect $f(L) \propto L^{-3}$
provided the correlation length is smaller than the system size.
In order to take into account all these possibilities
we have extrapolated our QMC data using 
$f(L) \propto L^{-\lambda}$ where $\lambda$ itself is a fit parameter.
Since the asymptotic behavior of $f(L)$ is only reached for
sufficiently large lattices 
we have checked that the omission of the data for small
lattices ($L = 4$ and $L = 6$) does not lead to different
conclusions concerning the existence or absence
of antiferromagnetic long range order.

The results of such a finite-size extrapolation are displayed in Fig. \ref{su6}
for $U = 6$ and $T = 0.3, 0.36$ and $0.4$.
While the curves for the two higher temperatures extrapolate to a value close to
zero indicating the absence of long-range order, the curve for $T = 0.3$
yields a finite value corresponding to
a finite sublattice magnetization.
This behavior is in agreement with the value $T_N \approx 0.31$ that we have 
extracted from the specific heat data. Preliminary results concerning the
magnetic structure factor for other values of $U$ have been published
elsewhere \cite{Sta99}. A more comprehensive presentation including
computational details can be found in \cite{Phd99}.

\begin{figure}[tb]
\centerline{\psfig{file=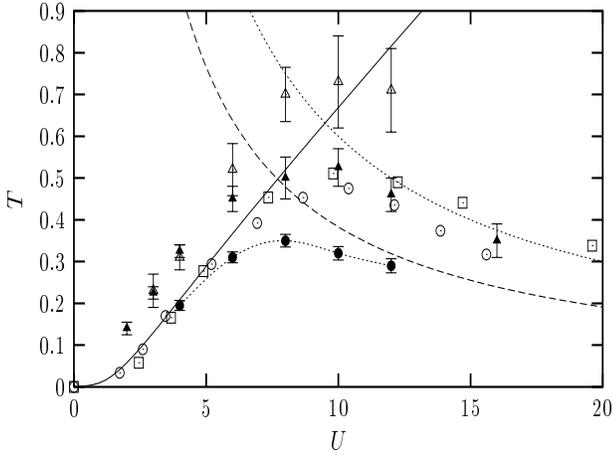,width=8cm,height=6cm,angle=0}}
\caption
{Magnetic phase diagram of the half-filled Hubbard model.
N\'eel temperature $T_N$ as function of $U$ from various approaches:
QMC, this work (filled circles; dots are meant as a guide to the eye only),
QMC \cite{Hir87} (open triangles), QMC \cite{Sca89} (filled triangles), 
DMFT \cite{Jar92} (open circles), DMFT \cite{Ulm95} (open squares),
modified Hartree-Fock theory \cite{Don9194} (solid curve),
Heisenberg limit from high temperature expansions, $T_N = 3.83 t^2 / U$
\cite{Rus74} (dashed curve),
Weiss molecular field theory, $T_N^{mf} = 6 t^2 / U$ (dotted curve).}
\label{phase}
\end{figure}

We summarize our results concerning the magnetic phase
transition in Fig. \ref{phase} where 
the N\'eel temperature of the 3D Hubbard model
obtained from different methods is displayed as function of $U$. 
For comparison the asymptotic behavior in the weak and strong coupling limit
is also shown. In Hartree Fock theory, applicable for small $U$,
$T_N$ is determined by the gap equation
\begin{equation}
\frac{2}{U} = \int {\rm d}\epsilon \,\frac{\rho(\epsilon)}{\epsilon} 
\,\tanh \frac{\epsilon}{2 T_N}
\label{hf}
\end{equation}
where $\rho(\epsilon)$ is the density of states.
As pointed out by van Dongen \cite{Don9194} the true asymptotic N\'eel temperature 
in the weak coupling limit is reduced by a factor $q \approx 0.282$ compared to
the solution of (\ref{hf}). We have included this reduction factor in the
curve shown in the figure.
In the strong coupling limit the Hubbard model can be mapped to the spin 
1/2 Heisenberg model
where the critical temperature is known from high temperature series \cite{Rus74}
and QMC simulations \cite{San98}
which yield $T_N = 3.83 t^2/ U$, compared to 
the Weiss molecular field result $T_N^{mf} = 6 t^2 / U$.
Our QMC results interpolate smoothly between weak and strong coupling 
asymptotics whereas the old QMC data of refs. \cite{Hir87} 
and \cite{Sca89} are clearly off in both limits.

Comparing our results with DMFT data is not straightforward since
in DMFT calculations mostly a Gaussian or a semi-elliptic density of states
is used which both differ from the non-interacting density of states
of the 3D Hubbard model. To convert energies we have expressed the hopping
matrix element $t$ (which is our energy unit) in terms of the
second moment of the density of states via
$t = \sqrt{\langle \epsilon^2\rangle/6}$. In practice this means that the 
energy data from ref. \cite{Jar92} have been multiplied by a factor $2 \sqrt 3$
and those from ref. \cite{Ulm95} by a factor of $2 \sqrt 6$.
In the weak coupling regime $U \lesssim 6$ there is good agreement between
the DMFT results
and our QMC data while for intermediate and large values
of $U$ the DMFT yields substantially higher values of $T_N$. This is not surprising since
in the limit $U \rightarrow \infty$ the DMFT reduces to the
Weiss molecular field theory of the Heisenberg model.
There are however significant discrepancies between the DMFT data from ref.
\cite{Jar92} and ref. \cite{Ulm95} for large $U$. This might be due to
the fact that in the first case a Gaussian and in the latter
a semi-elliptic density of states was used. 
It should also be noted that the DMFT data from \cite{Ulm95} approach 
the molecular field
limit from above while the curve of \cite{Jar92} and also our
QMC data lie always below their respective strong coupling asymptote.

\section{Metal insulator transition}

At a true metal insulator transition the DC conductivity $\sigma$ drops to zero
when some control parameter is changed across a critical value.
This is strictly speaking only possible at zero temperature whereas for $T > 0$
the conductivity remains always finite due to thermal activation.
Unfortunately, the DC conductivity -- although quite easily
accessible in experiment -- is very hard to obtain from QMC simulations, since the required
analytic continuation of the current-current correlation function from Matsubara
to real frequencies is numerically a very hard problem when the input data
are noisy. We have therefore based our considerations on calculations of
the electronic compressibility
$\kappa = \partial n /\partial \mu$ which can be obtained with high accuracy from
QMC simulations.  Although there is no simple relation between $\sigma$ and
$\kappa$ both are expected to be finite in the metallic
and exponentially small in the insulating regime.

DMFT calculations employing the iterated perturbation theory \cite{Geo96}
have revealed the following scenario for the Mott
Hubbard transition. In the fully frustrated Hubbard model
there exists a first-order metal-insulator transition line due to a coexistence
regime of metallic and insulating solutions at low temperatures. 
This first order line ends at a critical point,
reminiscent of an ordinary liquid gas transition \cite{Roz99}. 
Recent numerical work
has corroborated this scenario. In fact, the first order transition can
also be observed as a jump in the average number of doubly occupied
sites \cite{Roz99} and it is expected that the compressibility is 
discontinuous as well.

\begin{figure}[tb]
\centerline{\psfig{file=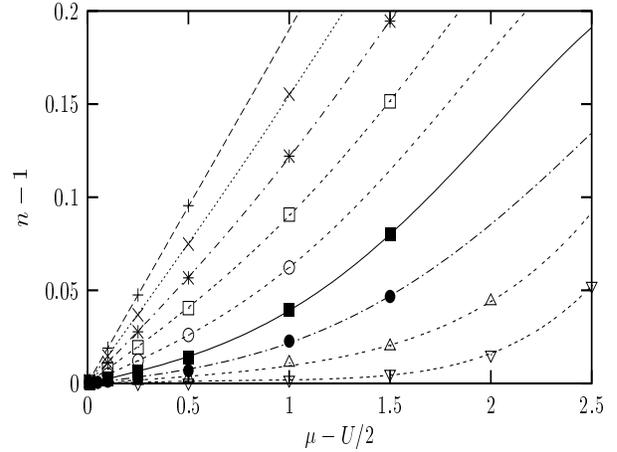,width=8cm,height=6cm,angle=0}}
\caption{Particle density $n$ as function of the chemical potential $\mu$ for
$U = 3,4,5,6,7,8,9,10,12$ (from top to bottom), $T = 0.45$ and $L = 4$.
The curves are polynomial fits.}
\label{nofmu}
\end{figure}

\begin{figure}[tb]
\centerline{\psfig{file=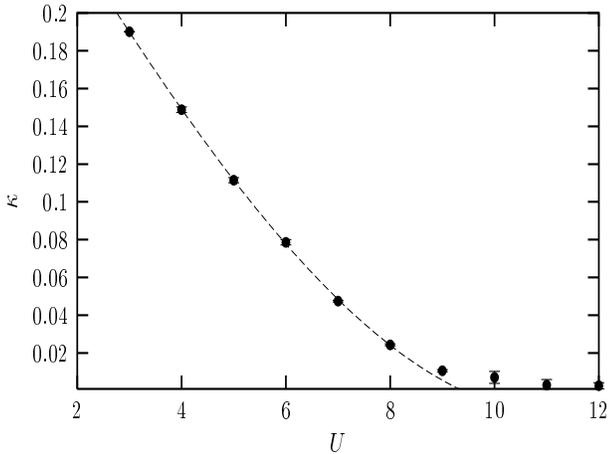,width=8cm,height=6cm,angle=0}}
\caption{Electronic compressibility $\kappa$ as function of $U$.
The curve is a polynomial fit for the range $3 < U < 8$.}
\label{kappaofu}
\end{figure}

Fig. \ref{nofmu} shows how the particle density $n$ depends on the chemical
potential $\mu$ for $T = 0.45$ which is slightly above the antiferromagnetic 
phase boundary. 
Due to particle-hole symmetry
$n - 1$ is an odd function of $\mu - U/2$. We have therefore 
used the polynomial $f(x) = a x + b x^3 + c x^5$ to fit the QMC data.
The coefficient $a$ yields the
compressibility $\kappa$ which is displayed as function of $U$ in Fig.
\ref{kappaofu}. There is no sign of a discontinuity as should be
observed in the case of a first order transition. Instead the compressibility 
decreases smoothly up to $U \approx 8$ and can be very accurately approximated by a
second order polynomial in this range, as indicated by the dashed curve in the figure.
Afterwards it turns over into an exponential-like tail that we associate with the
crossover regime where only thermal activation across the gap contributes
to the compressibility. Unfortunately our data are not precise enough to extract
the $U$ dependence of the gap.
The observation of a crossover regime instead of a true metal-insulator transition
is in agreement with the location of the critical point calculated within DMFT.
For a semi-elliptic density of states 
the critical values are $U_c \simeq 11.7$ and $T_c \simeq 0.127$ \cite{Roz99},
converted to our units
which is by a factor of more than two below the maximum N\'eel temperature i.e.
deep inside the antiferromagnetic phase.

\section{Conclusions}

We have performed QMC simulations of the half-filled 3D Hubbard model
on simple cubic lattices of up to $10^3$ sites.
Using finite-size scaling of the magnetic structure factor
it is shown that the low-temperature maximum of the specific heat
coincides with the antiferromagnetic phase transition.
The N\'eel temperature thus obtained interpolates smoothly between
the analytic solutions for weak and strong coupling but differs significantly
from the results of previous QMC simulations.
The shape of the specific heat curves close to the phase transition
is similar to the one obtained for the 3D spin 1/2 Heisenberg model
using high precision QMC simulations \cite{San98} indicating 
that both models are in the same universality class.
There is no indication of a first-order transition for small values of $U$
contrary to a conjecture established
in previous QMC simulations \cite{Hir87,Sca89}.

In order to investigate the transition from metallic to insulating behavior
in the paramagnetic phase above $T_N$ we have calculated the electronic
compressibility.
The transition appears to be broadened in accordance with the 
crossover scenario developed in the framework of the DMFT.
For $T = 0.45$ the crossover regime extends over the range
$9 \lesssim U \lesssim 11$ which is somewhat below what is obtained in DMFT
calculations \cite{Pru93}.

It would be very interesting to perform QMC simulations for a Hubbard
model where antiferromagnetism is suppressed by adding a next-nearest
neighbor hopping $t'$ in order to confirm the existence of the first
order transition line obtained within the DMFT. 
It is however to be feared that in this case the notorious minus sign
problem will prohibit efficient QMC simulations at low enough temperatures.
We estimate that the average sign of the fermion determinant 
behaves as $\propto \exp(-\beta t')$ which means that for temperatures
of the order of $t'$ the minus sign problem becomes severe and
therefore the development of improved QMC algorithms is needed to 
study this problem.

We thank P. van Dongen and P. Schwab for useful discussions.
This work was supported by the DFG as a part of the project {\em metal-insulator 
transition and magnetism in highly correlated transition-metal chalcogenids} (HO 
955/2).

\end{document}